# Tehran Stock Exchange Prediction Using Sentiment Analysis of Online Textual Opinions

Arezoo Hatefi Ghahfarrokhi, Mehrnoush Shamsfard

Faculty of Computer Engineering and Science, Shahid Beheshti University, Tehran, Iran

Arezuhatefi88@gmail.com, m-shams@sbu.ac.ir

## Abstract

In this paper, we investigate the impact of the social media data in predicting the Tehran Stock Exchange (TSE) variables for the first time. We consider the closing price and daily return of three different stocks for this investigation. We collected our social media data from Sahamyab.com/stocktwits for about three months. To extract information from online comments, we propose a hybrid sentiment analysis approach that combines lexicon-based and learning-based methods. Since lexicons that are available for the Persian language are not practical for sentiment analysis in the stock market domain, we built a particular sentiment lexicon for this domain. After designing and calculating daily sentiment indices using the sentiment of the comments, we examine their impact on the baseline models that only use historical market data and propose new predictor models using multi regression analysis. In addition to the sentiments, we also examine the comments volume and the users' reliabilities. We conclude that the predictability of various stocks in TSE is different depending on their attributes. Moreover, we indicate that for predicting the closing price only comments volume and for predicting the daily return both the volume and the sentiment of the comments could be useful. We demonstrate that Users' Trust coefficients have different behaviors toward the three stocks.

**Keywords**: stock market prediction, social media, sentiment analysis, textual analysis, natural language processing

## 1. Introduction

The stock market prediction has always been one of the demands of researchers and investors. If they can predict the future behavior of stock prices, they can quickly act based on this prediction and gain more profit. This desire has led them to many approaches for market analysis. Many theories have been suggested to explain stock market movements. Some of them focus on the underlying business behind a stock's price (fundamental analysis) (Greig, 1992; Mahmoud & Sakr, 2012), some focus on historical price movements (technical analysis) (Cervelló-Royo et al., 2015; Xiao & Enke, 2017), and some others focus on the human behavioral aspects of the market (behavioral finance) (Bollen et al., 2011; Gao, 2008; Keynes, 1936; Shleifer, 2000). One of the areas of behavioral finance revolves around the idea of the sentiment of the market participants. It means that in addition to historical prices, the current stock market is affected by the society's and investors' mood. Since rapid growth of Internet has led investors to share their opinions about the market in social media, forums, blogs, and etc., stock market prediction base on online sentiment tracking has drawn a lot of attention recently (Antweiler & Frank, 2004; Bollen et al., 2011; Nguyen et al., 2015; O'Hare et al., 2009; Nuno Oliveira et al., 2017; Wu et al., 2014). In this regard, microblogs are one of the most promising online resources for achieving investors' sentiment. Mao et al. found that Twitter has a strong predictive power even more than the predictive power of survey sentiment and news media analysis.

To the best of our knowledge, although there are several studies related to the prediction of the Tehran Stock Exchange (TSE) movement trends (Ahangar et al., 2010; Ebrahimpour et al., 2011; Fasanghari & Montazer, 2010; Zahedi & Rounaghi, 2015), none of them has considered sentiment. These studies are classified as technical analyses and they have only used historical prices and volume. In addition, these studies have used data mining techniques such as neural networks and genetic algorithms. It seems that this research is the first work aiming at investigating the effect of incorporating sentiment into the TSE prediction models. We investigated three symbols Vebmellat, Shabandar, and Khodro that belong to three different industries. We gathered users' comments about these stocks from Sahamyab.com/stocktwits for about three months. After extracting the sentiment of these comments by our proposed sentiment analysis method and making sentiment indices, we examined the impact of these indices on the baseline models. Since the reliability of the users affects the importance of their sentiments, we calculated a trust coefficient for each user based on his/her historical comments and incorporated them in several indices.

The sentiment classification techniques can be divided into the machine learning methods, lexicon-based methods, and hybrid methods. Machine learning methods apply the famous ML algorithms such as SVM and Naïve Bayes and use syntactic and linguistic features. These methods require labeled training data that is often difficult to obtain. The lexicon-based methods rely on generic or domain dependent lexicons or keywords. The hybrid approaches combine both methods and the lexicon plays mostly a key role.

Several studies have been conducted on sentiment analysis in the Persian Language. (Alimardani & Aghaei, 2015; Basiri et al., 2014; Saraee & Bagheri, 2013; Shams et al., 2012). Some of them have led to generating a lexicon which is either for the general domain or domains other than the stock market. In a comparison, we will show that generic lexicons are not appropriate for sentiment analysis in the stock market domain. In addition, Oliveira et al have reached a similar result in their study (N. Oliveira et al., 2016). In this paper, we propose a hybrid method for sentiment analysis in the stock market domain. First, we make a sentiment lexicon using the comments of this domain and then we use the lexical items of the lexicon as the classification features of the machine learning classification algorithms.

The rest of the paper is organized as follows. Section 2 provides a relevant literature, concentrating on some previous approaches of sentiment analysis for stock market prediction and sentiment analysis in the Persian language as well. Section 3 describes our dataset and our proposed method. Section 4 evaluates the results of the experiments. Section 5 discusses our results and concludes our contributions. Finally, future work is presented in section 6.

## 2. Related work

Using sentiment in financial markets was popularized in the early twentieth century with the introduction of Keynes Beauty Contest Analogy which argued that investors select the most beautiful (i.e., the most favorite) stock to invest in because they care the thought of other investors about that stock more than its real value (Keynes, 1936). Various investors use the concept of sentiment differently. As an example, when a contrarian investor recognizes that sentiment about the market is very negative, he may buy more stocks than usual because he believes great movements are coming to the market (Brown & Cliff, 2004).

At first various surveys such as National Association of Active Investment Managers (NAAIM) and American Association of Individual Investors (AAII) regular reports were used to evaluate investors and market sentiment. These surveys were used by many investors to understand the overall sentiment of the market, economy, and industries in order to make the necessary adjustments to their portfolios to take advantage of, or to protect themselves from, changes in market sentiment (Mian & Sankaraguruswamy, 2012). Despite these surveys popularity, they need a lot of resources and are expensive. In addition, they may face the problem of unreliable respondents, individual biases, social bias, and group thinking (Da et al., 2010; Singer, 2002).

In recent years, researchers have used a variety of methods to compute sentiment indicators using bulk online data. This approach is more appropriate than using surveys. First, computational analysis of the sentiment and the public mood is faster, more precise, and less costly than conducting large-scale surveys. Secondly, there is strong support for this claim that the sentiment obtained from this approach is a valid indicator of public opinion, as far as it is used to predict many socio-economic phenomena such as presidential elections (Burnap et al., 2016; Tumasjan et al., 2010; White, 2016), and Commercial sales (Choi & Varian, 2012; Liu et al., 2016; Mishne & Glance, 2006).

As far as we know, three distinct groups of online data sources have been used for financial forecasting. First, it has been shown that the content of the news media is an effective factor in the investor's sentiment and desire. Tetlock, for example, found that a high level of pessimism in the Wall Street Journal led to a decline in market returns on the following day (Tetlock, 2007).

Second, it has been indicated that web search query data is related to and even predictor of fluctuations in the stock market. The search volume of the stock names reveals the interests of investor, and therefore the high volume of searches for the name of a share reflects its price increase in the short term and the inversion of its price over the long term (Da et al., 2010). Also, the search volume has a strong correlation with the volume of traded shares. So that the peak of the search volume predicts the peak of the trading volume in one day or more (Bordino et al., 2012).

Finally, social media content has become an important data source for measuring the sentiment of society and investors. In an initial research, online stock message boards were used to predict market volatility and trading volume (Antweiler & Frank, 2004). In recent years, sentiment indicators extracted from social networks such as Facebook (Karabulut, 2013), Live Journal (Gilbert & Karahalios, 2010), and Twitter (Bollen et al., 2011) have been used to predict stock market fluctuations.

In addition, the content of the specific stock market social networks such as StockTwits.com or the website used in this research, Sahamyab.com, is reflective of most news related to the stock market, because their users are investors of financial markets and continually follow the economic and political news. Since the stock exchange of Iran is strongly influenced by the political and economic news because of specific circumstances of this country, a vast amount of news should be investigated, which we are approaching to by studying the comments of the users.

Antweiler and Frank (2004) investigated how Internet stock message boards are related to stock markets. They examined 1.5 million messages posted on two forums for 45 companies in the Dow Jones Industrial Average and the Dow Jones Internet Index and used naïve Bayes text classification to obtain sentiments of messages. They demonstrated a strong positive correlation between the messages of the online boards, trading volume, trading volatility, and a minor correlation between message board posts and price activity on the next day.

Das and Chen (2007) used a more formal approach to completing previous research on the use of sentiment analysis techniques for online message boards. In the previous research, a simple classification algorithm was used to assign the signal "buy," "sell," or "hold / neutral" to messages. Although these methods led to satisfactory results, the researchers sought to find a more automated and more robust classification method for categorizing messages in "bullish", "bearish", and "neutral" classes. Although the use of classification techniques requires more study, the importance of this study is to show that there is no significant correlation among sentiment and price movements of an individual share, but a positive correlation among the total sentiment of a set of stocks and their price changes.

Gu et al. (2006) and Zhang (2009) took a different approach to study stock message boards. In addition to the sentiment and volume of the messages, they also considered the reputation of the message senders. Both of these studies have shown that measuring the reputation of the message senders and then using them as a measure of whether or not to participate the messages in formulas or algorithms increases the accuracy of the model.

Wu et al. (2014) proposed a model for using stock related online posts to predict stock price volatility trends at both industry and individual stock company levels using sentiment analysis, SVM classifier and generalized autoregressive conditional heteroskedasticity (GARCH) modeling. They selected a typical financial website as an experimental platform to obtain a corpus of financial review data and examined both machine learning and lexicon-based approaches for sentiment classification and found that the precision of machine learning methods is greater than that of lexicon-based ones. Their results demonstrated that their GARCH-based SVM model performed better in predicting volatility trends at individual stock levels than at the industry level because the stock features effect on the predictive power of stock forum sentiment. Furthermore, there were significant differences in prediction accuracy between individual stocks regarding firm size and price-to-book features.

While all of the aforementioned researches have used online message boards, some other studies have used other forms of media such as blogs and microblogs to identify tendencies toward an individual stock or a group of stocks.

Bollen et al. (2011) analyzed the text content of daily Twitter feeds by two mood tracking tools, namely OpinionFinder and Google-Profile of Mood States (GPOMS). The former measures positive vs. negative mood and the latter measures mood in terms of 6 dimensions (Calm, Alert, Sure, Vital, Kind, and Happy). Afterward they used a Granger causality analysis and a Self-Organizing Fuzzy Neural Network to investigate the hypothesis that public mood states are predictive of changes in DJIA closing values. They indicated that adding some of the mood dimensions to the model could improve the accuracy of DJIA prediction. Their results showed 87.6% direction accuracy and a reduction of the Mean Average Percentage Error by more than 6%.

Sprenger et al. (2014) unlike previous researchers, focused on the stocks of the S & P 100 Index and collected tweets corresponding to these companies to study the correlation of sentiment expressed for a company on Twitter with the movements of its price and volume. The researchers tried to reduce a large number of unrelated tweets by only keeping tweets containing a dollar sign followed by the relevant ticker symbol. This method has been popularized by Stocktwits.com. This platform runs on top of the Twitter platform and allows StockTwits.com users to view Twitter content and vice versa. They used the Naïve Bayes classification method to assign a buy, sell, or hold signal to each tweet. Researchers found the bullishness of tweets to be associated with abnormal stock returns and message volume to predict next-day trading volume.

Unlike previous studies that used the overall moods or sentiments of documents, Nguyen et al. (2015) used the sentiment of specific topics of the company in prediction models. They proposed two methods for topic-level sentiment extraction. One is a JST-based method which relies on the existing topic model, the other is sentiment analysis at the feature level in which the features and sentiments are recognized by their proposed method. Although the average accuracy of their method for predicting price movement is 54.41%, it can predict price changes for a few stocks with an accuracy of over 60% and performs much better than other methods for the stocks that are difficult to predict with only past prices.

Nuno Oliveira et al. (2017) proposed a robust methodology to examine the value of microblogging data to forecast stock market variables: returns, volatility and trading volume of various indices and portfolios. They adopted a large Twitter dataset and used sentiment and attention indicators extracted from microblogs and survey indices (AAII and II, USMC and Sentix), diverse forms to daily aggregate these indicators, usage of a Kalman Filter to merge microblog and survey sources, a realistic rolling windows evaluation, several Machine Learning methods and the Diebold-Mariano test to assess if the sentiment and attention-based predictions are valuable in comparison with an autoregressive baseline. Their results showed that Twitter sentiment and posting volume were relevant for the forecasting of returns of S&P 500 index, portfolios of lower market capitalization and some industries. Additionally, KF sentiment was informative for the forecasting of returns. Moreover, Twitter and KF sentiment indicators were useful for the prediction of some survey sentiment indicators.

While researchers have investigated sentiment analysis of textual documents and web content for predicting stock prices and market activity in various world markets (Fisher et al., 2016), to the best of our knowledge, none of the previous studies on TSE prediction has considered the sentiment. These studies are categorized as technical analysis and they only used historical prices and volume as well as data mining techniques such as neural networks and genetic algorithms. In this research, we investigate the use of the sentiment analysis of the social media content in TSE prediction.

In general, there are few studies in the field of sentiment analysis and opinion mining for the Persian language. Shams et al. (2012) proposed an approach for building a lexicon for sentiment analysis in the Persian language. At first, they translated an existing English lexicon to Persian and then corrected errors of the previous step by using an iterative refinement approach. Finally, topic-based polar sets are achieved and each document is categorized into its related polarity with a classification algorithm. In addition, they presented an unsupervised LDA-based sentiment analysis method called LDASA. The proposed LDASA model modifies the LDA model by adding the polarity clues as sentiment features. According to their results, the proposed LDASA method significantly improve the accuracy compared to the base system which is only based on the term frequency of unigrams.

Saraee and Bagheri (2013) proposed a model for sentiment classification of Persian review documents. This model is based on lemmatization and feature selection and uses a Naïve Bayes classifier. In this research, the new Modified Mutual Information (MMI) feature has been introduced and its superiority to Mutual Information (MI) has been demonstrated.

Basiri et al. (2014) presented a lexicon-based approach for sentiment analysis in the Persian language. Their proposed framework includes modules such as spell correction and stemming for solving the problems of the Persian language processing. In addition, researchers have modified the SentiStrength tool, which is an English polarity lexicon, for Persian sentiment classification and translated its core into Persian. The results show that their proposed method on the mobile review dataset provides better F-score and less absolute error rather than machine learning methods.

Alimardani and Aghaei (2015) proposed a method for sentiment analysis in the Persian language by combining SentiWordNet and supervised algorithms. They made Persian SentiWordNet using a mapping between Persian WordNet and English SentiWordNet. Then this lexicon was used to weight features of supervised algorithms. They examined three different features for classification. The highest accuracy belonged to word frequency. In addition, they compared three machine learning algorithms: SVM, NB, and logistic regression and found that SVM and logistic regression performs better. The researchers used Kish hotels reviews to evaluate their model and reported 87% accuracy with SVM.

Hosseini et al. (2018) developed a manually annotated sentiment corpus, SentiPers, which covers formal and informal written contemporary Persian. SentiPers contains three different levels of annotation including document-level, sentence-level, and entity/aspect-level for Persian. The corpus contains more than 26000 sentences of users' opinions from digital product domain and benefits from special characteristics such as quantifying the positiveness or negativity of an opinion through assigning a number within a specific range to any given sentence.

It is noteworthy that besides the stock market domain, there is a wide range of research on sentiment analysis in other domains such as box office prediction (Du et al., 2014; Yu et al., 2012), business analytics (Coussement & Poel, 2009; Kang & Park, 2014), fraud detection (Goel & Uzlem, 2016), recommender system (Y.-M. Li & Shiu, 2012), medical domain (Noferesti & Shamsfard, 2015), and etc.

## 3. Material and methods

In this paper, first we propose a method to extract sentiment from online posts about TSE, and then we use this method to classify the posts into two classes: bullish and bearish. In the following, we examine the relations between extracted features from these posts such as their sentiments and their volume, and the changes in the values of the TSE. Finally, based on the results from previous experiments, we propose some prediction models for the under-investigation stocks. So, we can divide our approach to four steps:

- Collecting the data
- Building a sentiment lexicon for the stock market domain in Persian language and combining it with machine learning approaches for classifying the sentiment of the posts
- Designing and calculating a number of features including sentiment indicators, users' trust coefficients, and comments volume followed by investigating the relationship between these features and TSE variables
- Proposing prediction models for a number of stocks based on the previous results

## 3.1 Datasets

Our data includes the online textual comments and the TSE data. We collected textual comments from www.sahamyab.com/stocktwits which is the only social network for the stock market in Iran. Instead of 140-character limitation of Twitter, this website has a limitation of 2000 characters, so that users can attach their analysis and different news to their comments. Another feature of this site is the ability to tag the comments with bullish or bearish labels. To crawl Sahamyab.com we used Scraper, an extension for Chrome browser and extracted the comments of three stocks: Khodro[1], Shabandar[2], and Vebmellat[3] from May 1, 2016 to August 18, 2016. We extracted the text, time, label, user, and like count of the comments. TSE data was gathered from www.tsetmc.com.

However, some comments had sentiment labels, the number of these comments was insufficient for making a sentiment lexicon and training ML algorithms. In addition, the number of bearish comments was far fewer than the number of bullish comments. Therefore, we tagged the comments from May 1 to June 30 for Shabandar and Vebmellat whit bullish and bearish labels manually. The number of bullish comments was still more than the number of bearish ones in this new tagged dataset, so we separated a balanced dataset with an appropriate size from it. The number of bullish and bearish comments in these balanced and unbalanced datasets are presented in **Table *1***.

**Table 1**
The number of bullish and bearish posts in balanced and unbalanced datasets

| Class\Dataset | Unbalanced | Balanced |
|---|---|---|
| Bullish | 6248 | 2125 |
| Bearish | 2125 | 2125 |
| Total | 8373 | 4250 |

We select 2125 bullish comments for balanced dataset randomly from the total 6248 bullish ones in the unbalanced data.

## 3.2 Hybrid approach for sentiment analysis

The proposed model for sentiment analysis combines lexicon-based methods with machine learning approaches. Since many words have different polarity in the stock market domain than the general domain, using general lexicons for the stock market domain may lead to a low precision in classification. Therefore, we made a sentiment lexicon from a part of the data at first and then we calculated the sentiment scores for the other comments using this lexicon. Finally, we classified the comments to sentiment classes with machine learning methods. We used the abovementioned sentiment scores and the words of the lexicon as classification features. In the following, we explain these steps in detail.

---

[1] the stock of Iran Khodro company which is an Iranian multinational automaker
[2] the stock of Bandar Abbas oil refinery
[3] the stock of bank Mellat which is a private Iranian bank

### 3.2.1 Building Sentiment Lexicon

We used the unbalanced dataset for making the lexicon. This dataset included 2125 bearish and 6248 bullish comments.

*Preprocessing*

In addition to the common preprocessing steps which are done for any NLP task, we did some special preprocessing steps for our dataset:

- Removing the words beginning with @ (mentions) and # (tags)
- Substituting a single or a sequence of the sign "+" with the word "positive" and a single or a sequence of the sign "-" with the word "negative"
- Removing the redundant repetitive characters from the words
- Removing the colloquial stop words as well as the formal ones

Furthermore, for stemming and tokenization, we modified the tokenizer and stemmer tools of STeP-1 toolkit which performs normalization too (Shamsfard et al., 2010). This modification was made to process colloquial texts in addition to the formal texts. Many words in Persian have different writing forms that STeP-1 toolkit recognizes, processes and unifies them. For example, 'میرود' and 'می‌رود' and 'می رود ' are different forms of a formal word which means 'is going'. The colloquial forms of this word are 'میره' or 'می ره' or 'میره '. The modified version of Step-1 tokenizer can recognize and normalize the colloquial writing forms of the words too.

*Extracting the candidate words and phrases using word-level N-grams*

At this step, we extracted the appropriate candidates for inclusion in the sentiment lexicon. These candidates could be sentiment words which were attained from word-level unigrams or sentiment phrases such as nouns and modifiers which were obtained from word-level n-grams especially bigrams. For example, "green market" and "sales queue" in stock market domain have high sentimental significance.

One of the important tasks in sentiment analysis is managing negation in sentences because negation can change the polarity of the words and sentences. But, due to the complexities of the sentences in our dataset, using common methods for dealing with negation was really hard. Many of the sentences in our dataset were a combination of formal and colloquial words and/or grammar. This combination was an obstacle for parsing the sentence correctly. In addition, recognizing sentence boundaries was difficult due to the lack of punctuation marks in the comments. Using word-level bigrams helped to some extent solve this problem. For example the bigram 'بالا نمیره' which means 'does not go up' may have a negative polarity while 'بالا' (up) is positive.

*Stemming*

To reduce the number of candidates and remove redundancy from the lexicon, we tried to find the stems of the words and phrases using the stemmer of Step-1 toolkit.

The stemmer receives a single formal word and returns its stem(s), affix(es), lexical category, and other morphological features. In the cases that the stemmer could not find the stems of the colloquial words, these words were changed to the formal words with the same stem if there were any applicable transforming rules for them. Then, the resulting words were given to the stemmer. In the following, the transforming rules are presented. They were investigated in the same order as they are written. This order is based on being common in the dataset.

1. For making plural words in the Persian language, we add "ها" (/hâ/) or "ان" (/ân/) to the end of the single words. But, in the colloquial language, both of these morphemes are changed to "ا" (/â/). So, to reach the single form of the word we must delete "ا" (/â/) from the end of the word.
2. In the Persian language, "را" (/râ/) is a preposition that always follows the direct object of the sentence. But, in the colloquial language this word is changed to "و" (/o/) that adheres to the end of the object. In this case, we must eliminate this letter from the end of the word to reach a formal word.
3. In the Persian Language, "مان" (/mân/, our), "تان" (/tân/, your), and "شان" (/ʃân/, their) are continuous plural personal pronouns that show possession. These pronouns are substituted by "مون" (/mun/), "تون" (/tun/), and "شون" (/ʃun/) in the colloquial language. So, to find the word stem, we must delete these morphemes from the end of the word.
4. In the colloquial Persian language, that part of the verb which expresses the person and the number of the verb is changed in some verb tenses. These cases are presented in the **Table 4**. In these cases we must change this part of the verb to reach its formal form.

**Table 2**
Changes in the verb form in the colloquial Persian Language

| Tense | Person | Number | Change | Example |
|---|---|---|---|---|
| Past simple, Past continuous, Past perfect | 2nd | plural | "ید"(/id/) to "ین"(/in/) | "رفتین" to "رفتید" |
| Past simple, Past continuous, Past perfect | 3rd | plural | "ند"(/ænd/) to "ن"(/æn/) | "رفتن" to "رفتند" |
| Present simple, Present subjunctive | 2nd | plural | "ید"(/id/) to "ین"(/in/) | "می‌خورین" to "می‌خورید" |
| Present simple, Present subjunctive | 3rd | plural | "ند"(/ænd/) to "ن"(/æn/) | "می‌خورن" to "می‌خورند" |
| Present simple, Present subjunctive | 3rd | singular | "د"(/æd/) to "ه"(/e/) | "می‌خوره" to "می‌خورد" |

Finally, if the stemming process didn't find the stem of the word, the word itself was considered as its stem. In this way, it was possible for the colloquial words to be included in the final lexicon besides the formal words.

Since a negative verb can change the sentiment of the whole sentence, we made difference between the negative and the positive verbs. So, we added the roots of the negative verbs to the lexicon in a special format to show their negation.

To stem the bigrams, at first, each phrase was given to the stemmer as a whole. In the case of failure, we tried to find the stems of the constituents of the phrase and then put them together to generate the stem of the primitive bigram.

*Selecting the items of the lexicon and calculating their scores*
After calculating document frequency (DF) for the stems we founded in the previous step by assuming each post as a document, we selected the terms whose DF were more than a threshold. In this way, many of the misspellings would be removed from the lexicon.

Afterward, we used pointwise mutual information (PMI) statistic for measuring the dependency of the lexical items to each class. PMI is a measure of how much the actual probability of a particular co-occurrence of events $p(x, y)$ differs from what we would expect it to be on the basis of the probabilities of the individual events and the assumption of independence $p(x)p(y)$. The formula of PMI is as follows:

$$PMI(x, y) = \log \frac{p(x,y)}{p(x)p(y)} \tag{1}$$

Where x and y are the events and $p(x, y)$ is the probability of their co-occurrence. PMI will be highly positive if *x* and *y* are strongly dependent, highly negative if they are complementary, and nearly zero if there is no significant relationship between them. We defined the sentiment score of the lexical items in a way that included the PMI for both classes:

$$S_{PMI}(l) = PMI(l, bullish) - PMI(l, bearish) \tag{2}$$

Where *l* is a lexical item, bullish means the bullish class, and bearish means the bearish class. We calculated $PMI(l, bullish)$ and $PMI(l, bearish)$ by the use of the contingency table that is represented in (Saraee & Bagheri, 2013).

This way, the sentiment score reflected the sentiment orientation of the word. We just added those bigrams to the lexicon whose scores were larger than the sum of the scores of their constituents.

### 3.2.2 Building classification model

After making the sentiment lexicon, we tried to train a classifier for categorizing comments to bullish and bearish classes. At first, we made the feature vectors for the training data. The features were:

- Lexicon items (both unigrams and bigrams): the values were computed by the multiplication of the item frequency in the comment by its sentiment score. For counting the number of unigrams in the comment, we ignored those occurrences that were part of a bigram.
- Comment score: since the comments in the dataset had different length, we incorporated the comment length to the comment score. So, the comment score was computed by dividing the sum of the scores of the unigrams and bigrams of the comment to the comment length.

$$Score_{comment} = \frac{\sum Score_{uni} + \sum Score_{bi}}{Length_{comment}} \tag{3}$$

After preparing these vectors, we examined popular classification algorithms including SVM, Naïve Bayes, and Decision Tree for labeling each comment as bullish or bearish.

## 3.3 Investigating the relation between the TSE and sentiments

In this research, we wanted to investigate the relation between users' sentiments and TSE. To fulfill this goal, we built some sentiment indicators based on the comments and their sentiments. Then, we tried to build some predictor models by incorporating these sentiment indices to the models which was only using the stock market data for prediction.

Users are not always honest when they are commenting about stocks in social media. Sometimes, they advertise a bad stock or degrade a profitable one for some personal reasons. This issue is a barrier for obtaining the true sentiment about stocks. Therefore, we calculated users' reliability based on the history of their comments (if there were any) and the real stock movements in the period of June 9, to June 30, 2016. Then, we used these user reliability values in some of our sentiment indicators.

We will explain sentiment indices, trust coefficients, and predictor models in the following subsections.

### 3.3.1 Computing Users' Trust Coefficients

We computed the trust coefficient for stock s and user u by formula 4:

$$TC(u,s) = \frac{\sum_{i=1}^{N}(Correct(u,s,i)/Total(u,s,i))*(Total(s,i)/Correct(s,i))}{N} \tag{4}$$

Where Correct(u,s,i) is the number of correct comments and Total(u,s,i) is the number of total comments of user u about stock s in day i. While Correct(s,i) is the number of correct comments and Total(s,i) is the number of total comments of all users about stock s in day i. N is the number of days in the period whose comments we used for calculating the coefficients.

A comment about a stock was considered a correct comment if it predicted the change direction (up or down) of the price of that stock for the day after the publishing date of the comment correctly, considering the stock's real changes. If the number of the total comments of a user was greater than the average number of the comments of all users, we used this coefficient for him/her, otherwise, we used ½.

### 3.3.2 Computing Daily Sentiment Indices

We defined four daily sentiment indices for each stock as follows:

$$Sentiment\ Index_1(s,i) = \frac{Count(s_{bullish},i)}{Count(s_{bullish},i)+Count(s_{bearish},i)} \quad (5)$$

Where $Count(s_{bullish},i)$ is the number of bullish comments and $Count(s_{bearish},i)$ is the number of bearish comments for stock s in day i. We named this index the "Bullishness Index". It was used by Mao et al. (2011) as well with the name of Twitter Investor Sentiment (TIS).

To consider the effect of the users' trust coefficients on prediction, we introduced the second index as below

$$Sentiment\ Index_2(s,i) = \frac{\sum TC(optimistic\ users)}{\sum TC(optimistic\ users)+\sum TC(pessimistic\ users)} \quad (6)$$

Where TC() refers to the trust coefficient. Optimistic users are those users who tag their comments with the bullish label and pessimistic users are similarly defined. We called this index the "Bullishness With Trust Index". For calculating this index for each stock, we used the

Since we calculated these coefficients for each stock separately, so here we used the special coefficients of stocks.

For the third index, we used not only the direction prediction but also the sentiment scores of the comments to employ the extent of their optimism or pessimism.

$$Sentiment\ Index_3(s,i) = \frac{Score(s_{bullish},i)}{Score(s_{bullish},i)+Score(s_{bearish},i)} \quad (7)$$

We called this index the "Bullishness Score Index" in which Score($s_{bullish}$,i) is the sum of scores of bullish comments and Score($s_{bearish}$,i) is the sum of scores of bearish comments for stock s in day i.

Finally, in the fourth index we combined the comment scores with trust coefficients:

$$Sentiment\ Index_4(s,i) = \frac{\sum_{j=1}^{Count(s_{bullish},i)} Score(s_{bullish,j},i) * TC(u_j,s)}{\sum_{j=1}^{Count(s_{bullish},i)} Score(s_{bullish,j},i) * TC(u_j,s) + \sum_{j=1}^{Count(s_{bearish},i)} Score(s_{bearish,j},i) * TC(u_j,s)} \quad (8)$$

We named this index "Bullishness Score With Trust Index".

In addition to these indices, we extracted two more features from microblogging data for each stock:

- The total number of the comments in each day
- The number of the comments + the number of their likes in each day which is a criterion of the attention to that stock.

Among the variables of the stock market, we chose the closing price and the stock return which according to (Ranco et al., 2015) is calculated as below

$$Return(s, i) = \frac{Close(s,i) - Close(s, i-1)}{Close(s, i-1)} \quad (9)$$

Where $Close(s, i)$ is the closing price of stock s at the end of day i.

### 3.3.3  Predictor Models

At first, we made autoregressive baseline models $M_0$ for the closing price and the return value of each under investigation stock. Then, we tried to build the more advanced models $M_1$ by incorporating the features we extracted from Sahamyab.com to the former models.

In the literature, two approaches have been considered for making predictor models: classification approaches which only predict the direction of the changes (up and down) and the regression models. We used the latter method in this research. We will describe the steps of creating $M_1$ models in the following parts.

**Autocorrelation analysis**

First, we performed an autocorrelation analysis for each feature of the stocks to find the lags that the feature had an autocorrelation relation with. Afterward, we created the autoregressive baseline models based on these lags. Among these lags, we kept those lags that had statistically significant coefficients in the model and removed the others. The form of the baseline models is as follows:

$$M_0 : Y(t) = \sum_{i=1}^{n} \alpha_i * Y_{t-i} + \varepsilon_t \quad (10)$$

In general, after building a time series model, the residuals should be a white noise. So, there shouldn't be any autocorrelation relation between them. Otherwise, the time series model is wrong. Therefore, we performed an autocorrelation analysis for the model residuals to be sure of the accuracy of the model.

**Cross-correlation analysis**

For making $M_1$ models, we needed to find the lags of independent variables (features extracted from Sahamyab.com) that were correlated with the dependent variable (stock features). We used cross-correlation analysis for discovering these relations.

**Granger Causality Test**

Since correlation doesn't imply causation, we performed Granger causality test for the correlations we had found in the previous step. As an example, if there is a correlation between feature A of stock s and feature B extracted from Sahamyab.com in some lags and Granger causality test proves causation from B to A in these lags, we incorporate feature B to $M_1$ model of feature A.

**Multiple regression**

After discovering causality relations in the previous step, we created a multiple regression model for each stock feature by adding the correlated independent variables to $M_0$ models. We chose the variables for the insertion to the M0 model based on the descending order of the strength of the correlation relations. If insertion of a variable to $M_0$ resulted in insignificant coefficients for previously added variables, we undid this insertion. Furthermore, at each step, we included all the significant lags (P-value < 0.05) of the independent variable which we found in the previous steps and then we just kept those lags that had statistically significant coefficients. We used 90% of our data for building these models and the other 10% for testing them. The form of $M_1$ models would be as follows:

$$M_1 : Y(t) = \alpha + \sum_{i=1}^{n} \beta_i * Y_{t-i} + \sum_{i=1}^{m} \gamma_i * x_{t-i} + \varepsilon_t \tag{11}$$

### 3.3.4 Comparing Models

After making $M_0$ and $M_1$ for each stock feature, we used 10% data for comparing them. Our evaluation measures were:

- **MAPE** (Mean Absolute Percentage Error):

$$MAPE = \frac{\sum_{i}^{n} \left| \frac{y_i - \hat{y}_i}{y_i} \right|}{n} * 100 \tag{12}$$

In which $\hat{y}_i$ is the predicted value and $y_i$ is the real value of dependent variable y.

- **DA** (Direction Accuracy):

$$\sum (\widehat{y_{t+1,\iota}} - \widehat{y_{t,\iota}}) * (y_{t+1,i} - y_{t,i}) > 0 \tag{13}$$

It means that if the direction prediction for the changes of variable y in two sequential days is the same as the direction of real changes in these days, that prediction is accurate.

## 4 Results

Evaluation of our approach included two parts: the evaluation of the suggested sentiment analysis method for the stock market domain and the evaluation of the proposed prediction models for the stock features. For getting reliable results, all the numbers that are reported in the tables were obtained by the use of the 10-fold cross-validation method.

### 4.1 Lexicon selection

For selecting the best lexicon, a number of comments were classified as bullish and bearish by the means of two lexicons that were described in section 3.1 as the features of the classifiers. The used classifiers were naïve bays and decision tree classifiers. In all experiments, the lexicon built from the balanced dataset leaded to better results. So, we chose this lexicon as the classification features for the rest of the experiments.

### 4.2 Selection of the best classification algorithm for sentiment classification

Naïve bays, Decision tree, SVM, and bagging classification algorithms were examined to find the best one for our sentiment classification task. **Table 3** shows the 10-fold cross-validation results for these four algorithms and it is seen that the bagging algorithm is the most appropriate one according to F-measure.

**Table 3**
Comparison of four classifiers for sentiment classification in stock market domain

| Algorithm | Accuracy | Recall | F-measure |
|---|---|---|---|
| NB | 75.8% | 75.5% | 75.4% |
| DT | 84.8% | 84.8% | 84.8% |
| SVM | 85.7% | 84.6% | 84.6% |
| Bagging(J48) | 85.5% | 85.3% | **85.3%** |

## 4.3 Effect of "Score" feature in sentiment classification

To investigate the effect of sentiment score feature that was attached at the end of the feature vectors for classification, one experiment with this feature and one experiment without it were performed. **Table *4*** shows that this feature has a great effect on the accuracy of our sentiment classification task.

**Table 4**
Results of two classification experiments for examining the effect of score feature

| Experiment (Bagging) | Accuracy | Recall | F-measure |
|---|---|---|---|
| With Score Feature | **85.5%** | **85.3%** | **85.3%** |
| Without Score Feature | 66.1% | 65.5% | 65.1% |

## 4.4 Evaluation of the proposed sentiment analysis algorithm for the test data

Toward evaluating our proposed sentiment analysis algorithm, we applied this method to a set of 183 comments including 97 bearish and 86 bullish comments. These comments were not used for building lexicons and training classifiers. The results are shown in **Table *5***.

**Table 5**
Results of our sentiment analysis algorithm for the test data

| Class | Accuracy | Recall | F-measure |
|---|---|---|---|
| Bullish | 75% | 76.7% | 75.9% |
| Bearish | 78.9% | 77.3% | 78.1% |
| Total | 77.1% | 77% | 77.1% |

## 4.5 Comparison of our sentiment analysis approach with other work in Persian language

Among sentiment analysis methods for the Persian language, the most similar one to our approach is (Alimardani & Aghaei, 2015). In this research, the data was collected from hellokish.com and included some reviews about hotels in Kish Island. In order to make a lexicon they extracted frequent words from these reviews, translated them into English, and finally obtained their semantic orientations using SentiWordNet. Since SentiWordNet is for the general domain, it was not very applicable for the stock market domain. Therefore, we used the comments from the stock market domain to create a domain-specific lexicon and calculated the scores using the PMI measure. Furthermore, unlike our approach, researchers in (Alimardani & Aghaei, 2015) performed preprocessing steps such as editing half-space, transforming colloquial words to formal words, and spell checking manually. Another difference between our approach and this work is the attachment of the score feature to the feature vector of the classifier that improves the classification accuracy.

To show the superiority of domain-specific lexicon over open-domain lexicon for our sentiment analysis task, we performed two experiments with our test dataset. First, we classified the comments using our lexicon and then classified them using the sentiment lexicon of University of Tehran[4] as classification features.

---

[4] http://ece.ut.ac.ir/404

**Table 6** represents that using domain-specific lexicon for our sentiment analysis task resulted in better performance.

**Table 6**
Comparison of open-domain and domain-specific lexicons for sentiment analysis in the stock market domain

| lexicon | Class | Accuracy | Recall | F-measure | CC1 | CC2 |
|---|---|---|---|---|---|---|
| Domain-specific lexicon | Bullish | **72.8%** | 77.9% | **75.2%** | **74.3%** | **75.5%** |
|  | Bearish | **76.7%** | 71.1% | 73.7% |  |  |
| Open-domain lexicon (UT) | Bullish | 51.7% | **88.3%** | 65.2% | 54% | 55% |
|  | Bearish | 63.8% | 23.7% | 34.5% |  |  |

CC1 is the percentage of comments that were classified correctly with considering neutral comments and CC2 is the percentage of comments that were classified correctly without considering neutral comments

For comparing our approach with the work of Saraee and Bagheri (2013) we applied both models to our test dataset. The results are presented in **Table 7**. These results show that our approach outperforms their proposed model.

**Table 7**
The comparison of our approach with the proposed approach in (Saraee & Bagheri, 2013)

| model | class | Accuracy | Recall | F-measure |
|---|---|---|---|---|
| Our approach | Bullish | **73.6%** | **74.4%** | 74% |
|  | Bearish | **77.1%** | **76.3%** | 76.7% |
| Proposed model in (Saraee & Bagheri, 2013) | Bullish | 65.6% | 70.9% | **68.2%** |
|  | Bearish | 72.2% | 67% | **69.5%** |

### 4.6 Prediction models

In this section, we present and compare the baseline models and the proposed prediction models for closing price and daily return features of investigated stocks. We used 1-step ahead prediction method for comparing these models and we evaluated prediction accuracy by means of MAPE and DA measures.

#### 4.6.1 Bank Mellat (Vebmellat) stock

In this section the M0 and M1 models for closing price and daily return features of Vebmellat stock are presented and compared.

**Closing price feature**

By performing the steps of building $M_0$ and $M_1$ models that are mentioned in section 3.3.3, the baseline model and the proposed model for closing price feature of Vebmellat stock would be as follows:

$$M_0: close_t = 0.9868 * close_{t-1} \tag{14}$$

$$M_1: close_t = 0.9566 * close_{t-1} + 0.2856 * countWithLikes_{t-1} \tag{15}$$

In **Table 7**, these two models have been compared with respect to the predicted values for the test data. According to the table, $M_1$ model has improved direction accuracy by 40% and has decreased MAPE by 0.26%. These promising results indicate the superiority of $M_1$ to $M_0$. According to these results, employing the number of comments plus the number of their likes into $M_0$ model can drastically improve its performance.

**Table 8**
Comparison of $M_0$ and $M_1$ models for the closing price feature of Vebmellat stock with respect to predicted values for test data

| Model | MAPE | DA |
|---|---|---|
| M0 | 0.97% | 20% |
| M1 | 0.71% | 60% |

As we described in section 3.3.3, one of the necessary steps for developing $M_0$ model to $M_1$ model is to find cross-correlation relations between closing price feature and extracted features from our data. In **Figure 1**, the cross-correlation between closing price feature of Vebmellat and our independent variables are shown for different lags. The closing price feature is significantly correlated with independent features when the correlation value is upper or lower than the positive or negative correlation bounds which are shown with blue dashed lines in the figures. For example, countWithLikes$_{t+k}$ (the number of comments plus the number of their likes at time t+k) is significantly positively correlated with closing price$_t$ when k = -1, -2, -3. It means that countWithLikes independent variable leads closing price variable at lags 1, 2, 3. Since correlation doesn't imply causation, the lags 2 and 3 removed in next steps and just the first lag was incorporated in the final model.

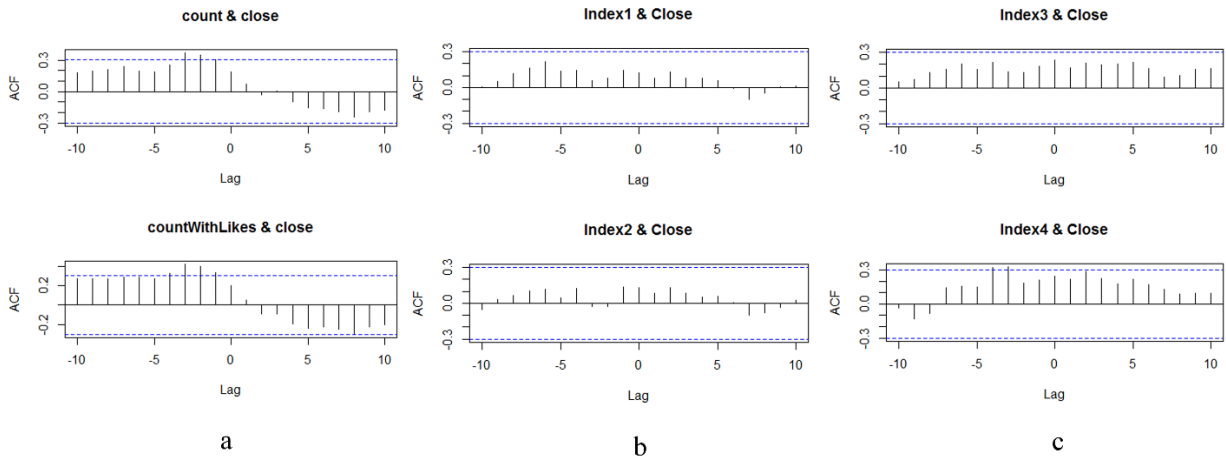

a  b  c

**Figure 1** The cross-correlation of the closing price of Vebmellat stock with comments count and sentiment indices

**Figure 2** represents predicted values for test data. The figure shows that predicted values by $M_1$ model are closer to real values in most cases.

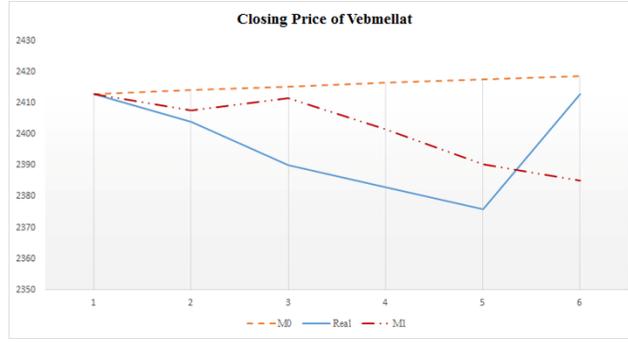

**Figure 2** Predicted values for the closing price of Vebmellat for test data by two models $M_0$ and $M_1$

**Daily return**

By performing the steps of building $M_0$ and $M_1$ models that are mentioned in section 3.3.3, the baseline model and the proposed model for daily return feature of Vebmellat stock would be as follows:

$$M_0 : return = 0.3821 * return_{t-1} \qquad (16)$$

$$M_1 : return_t = 0.0002 * countWithLikes_{t-1} - 0.0014 * Index4_{t-5} \qquad (17)$$

In **Table 7**, these two models have been compared with respect to the predicted values for the test data. According to the table, although $M_1$ model has not affected direction accuracy, it has decreased MAPE by 47.6%. As can be seen in the table, the direction accuracy of $M_0$ model is as high as the direction accuracy of $M_1$ model but its MAPE is very high.

**Table 9**

Comparison of $M_0$ and $M_1$ models for daily return feature of Vebmellat stock with respect to predicted values for test data

| Model | MAPE | DA |
|---|---|---|
| M0 | 143.3% | 80% |
| M1 | 95.7% | 80% |

In **Figure 1**, the cross-correlation between daily return feature of Vebmellat stock and our independent variables are shown for different lags. As can be seen in the figure, the countWithLikes (the number of comments plus the number of their likes) variable is significantly positively correlated with daily return at lags 0 and 1 and $Index_4$ is significantly negatively correlated with daily return at lag 5.

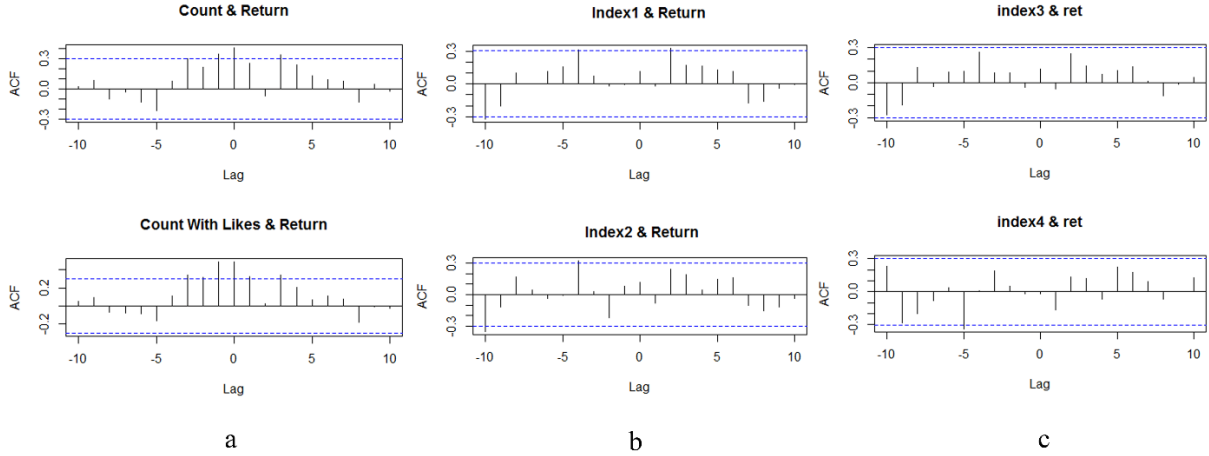

**Figure 3** The cross-correlation of daily return of Vebmellat stock with comments count and sentiment indices

**Figure** *2* represents predicted values for test data by two models $M_0$ and $M_1$.

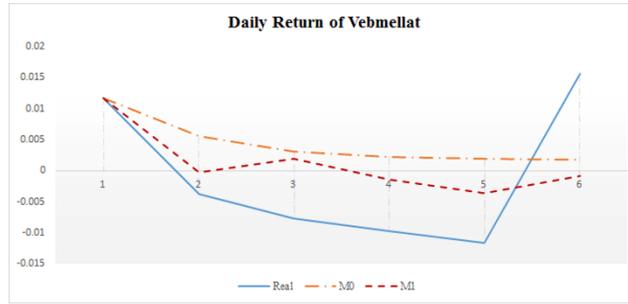

**Figure 4** Predicted values for the daily return of Vebmellat for test data

### 4.6.2 Bandar Abbas refinery (Shabandar) stock

In this section the M0 and M1 models for closing price and daily return features of Shabandar stock are presented and compared.

**Closing price**

The baseline model and the proposed model for the closing price feature of Shabandar stock would be as follows:

$$M_0 : close_t = 0.9607 * close_{t-1} \qquad (18)$$

$$M_1 : close_t = 0.9882 * close_{t-1} + 0.4917 * countWithLikes_{t-5} \qquad (19)$$

In **Table 7** these two models have been compared with respect to predicted values for test data. In $M_1$ model, direction accuracy has improved by 43% and MAPE has decreased by 0.47%. These promising results indicate the superiority of $M_1$ to $M_0$ in predicting the closing price feature of Shabandar stock.

**Table 10**
Comparison of $M_0$ and $M_1$ models for the closing price of Shabandar stock with respect to predicted values for test data

| Model | MAPE | DA |
|-------|------|-----|
| M0 | 1.6% | 28% |
| M1 | 1.13% | 71% |

**Figure 1** represents predicted values for test data by two models $M_0$ and $M_1$. As can be seen, the direction of the movement of the predicted values by $M_1$ model is alike the direction of the movement of the real values in most parts of the figure. However, the predicted values with the baseline model are constantly declining. In addition, the MAPE is small for both models which means the predicted values by both models are close to the real values.

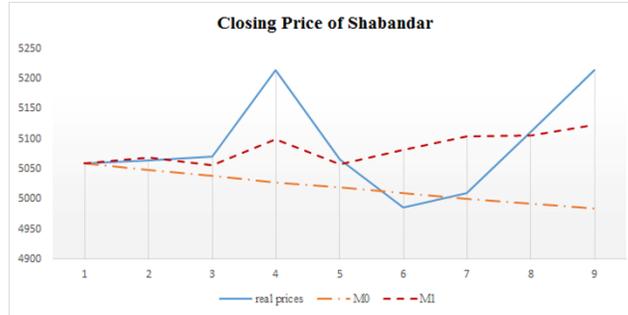

**Figure 5** Predicted values for the closing price of Shabandar for test data

**Daily return**

The baseline model and the proposed model for the daily return feature of Shabandar stock would be as follows:

$$M_0 : return_t = 0.4882 * return_{t-1} \qquad (20)$$

$$M_1 : return_t = 0.4934 * return_{t-1} - 0.0001 * countWithLikes_{t-5} - 0.0322 * Index3_{t-2} \qquad (21)$$

In **Table 7** these two models have been compared with respect to predicted values for test data. According to the table, although direction accuracy has decreased in $M_1$, there exists a big reduction in MAPE.

**Table 11**
Comparison of $M_0$ and $M_1$ models of the daily return for Shabandar stock due to predicted values for test data

| Model | MAPE | Direction Accuracy |
|-------|------|--------------------|
| M0 | 89.9% | 87.5% |
| M1 | 38.2% | 75% |

**Figure 1** illustrates predicted values for test data by two models $M_0$ and $M_1$.

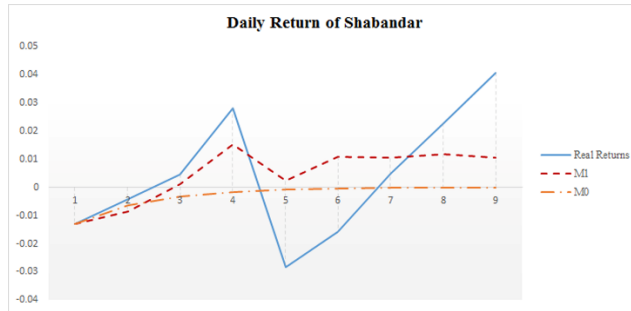

**Figure 6** Predicted values for the daily return of Shabandar for test data

### 4.6.3 Iran Khodro (Khodro)

**Closing Price**

As can be seen in **Figure 1**, none of our independent features was significantly correlated with the closing price feature of Khodro stock. Moreover, Granger causality test indicated no statistically significant causality relation between the closing price of Khodro stock and our independent features. Hence, making the $M_1$ model and comparing it to the $M_0$ model was not meaningful.

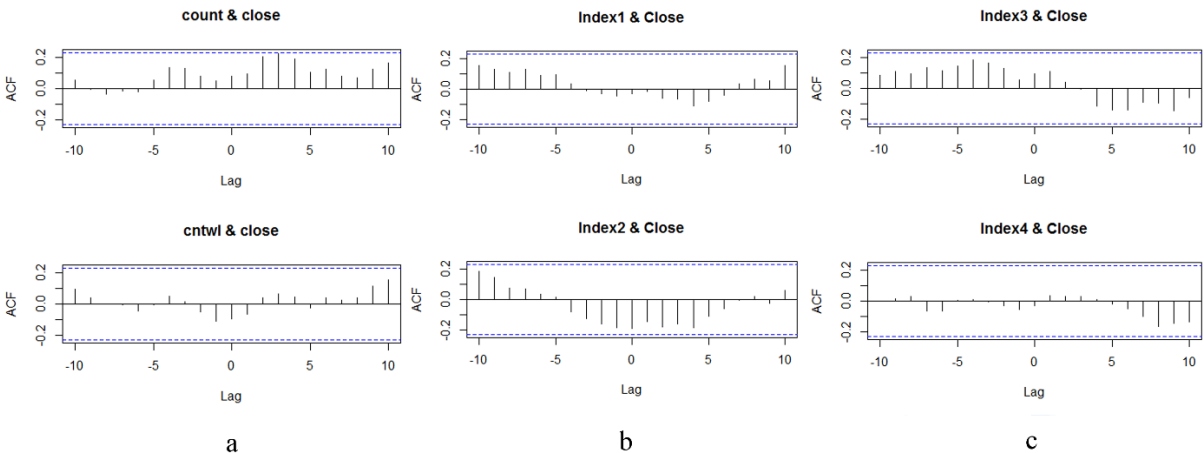

**Figure 7** The cross-correlation of the closing price of Khodro stock with comments count and sentiment indices

**Daily Return**

As can be seen in **Figure 8**, daily return of Khodro was significantly negatively correlated with the count of comments in lags 5 and 6. However, performing Granger test indicated no causality for these lags. Thus, making the $M_1$ model and comparing it to $M_0$ is not meaningful for daily return of Khodro stock..

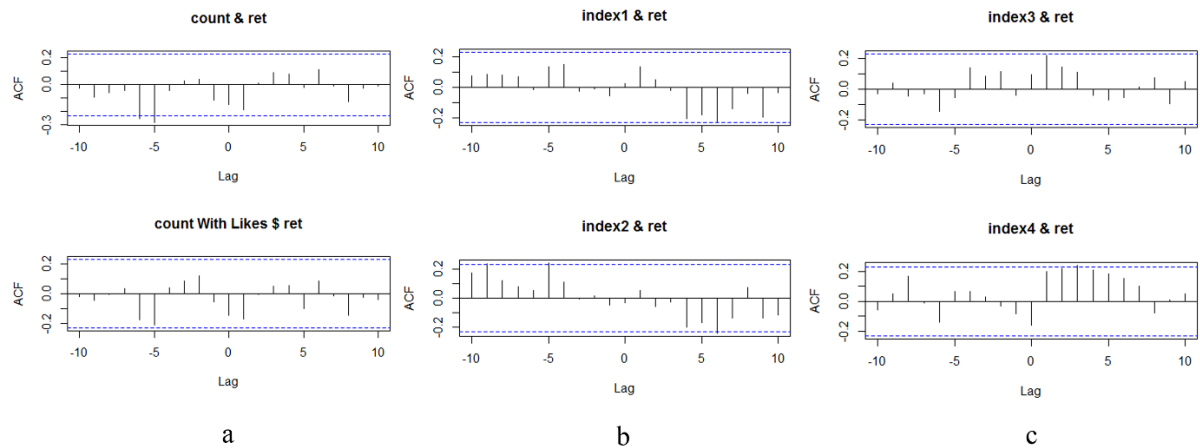

**Figure 8** The cross-correlation of the daily return of Khodro stock with comments count and sentiment indices

## 5 Conclusions

The task of stock market prediction is intrinsically very difficult. There are lots of internal or external factors that have linear or non-linear relations with variables of these markets. These variables can affect the value and movement of the market variables. In addition, the political and economic conditions of Iran, particularly in recent years, have led to high instability in TSE and have made its prediction even harder.

In (Das & Chen, 2007) researchers reported that although the sentiment is not useful for predicting the movement of individual stocks, it provides a valuable prediction method when it is applied to a group of stocks or to an index based on a group of stocks. In this paper, we investigated three stocks of TSE individually and we concluded that although the features that we extracted from Sahamyab were not strongly correlated with the features of the investigated stocks, they could improve the baseline prediction models in some cases. Accordingly, it can be said that the efficiency of using sentiment in prediction models is different depending on the attributes of the stocks. For example, Vebmellat stock and Shabandar stock have better fundamental analysis than Khodro stock and we saw that sentiment was not beneficial in prediction models of Khodro stock. This result is consistent with (Wu et al., 2014) in which researchers reported that prediction at individual stock level will have preferable results than prediction in industry level because features affect the predictive power of stock forum sentiment. In addition, the researchers in (Q. Li et al., 2014) found that the impact of media on firms varies according to firm characteristics and article content that confirms our conclusion too.

In addition, we found that adding the number of comments to the number of their likes for Vebmellat and Shabandar resulted in stronger correlation with two under investigation stock features. So, we concluded that the sum of comments with their likes is a measure of investors' attention to the stock.

Moreover, we showed that although the closing price feature was correlated with sentiment indices in some lags, these indices didn't enhance prediction performance for this feature and $M_1$ models for the closing price were based on daily number of comments. However, for making $M_1$ models for the daily return of Vebmellat and Shabandar we used the number of comments and the sentiment indices as well.

Trust coefficients showed different behavior toward three investigated stocks. For example, in the case of Khodro stock, $index_1$ and $index_3$ had stronger correlations with stock features than $index_2$ and $index_4$ had, with the knowledge that $index_1$ and $index_3$ are changed to $index_2$ and $index_4$ by incorporating trust coefficients. The reason was that, in the investigated time period in this research, Khordo stock was one of the most attractive stocks for investors and a lot of impermanent users commented for this stock in social media. While, the data that we used for calculating trust coefficients was limited. So, trust coefficients were not very reliable for Khodro stock. However, Vebmellat stock was a more stable stock and its investors were almost constant. Accordingly, the trust coefficients calculated for Vebmellat were more reliable. For this stock, $index_2$ and $index_4$ had storonger correlations with stock features than $index_1$ and $index_3$ had. The best feature for daily return of Vebmellat stock was $index_4$.

# 6  Future works

One of the weaknesses of this research is that for investigating the relationship between investors' sentiment and TSE prices and values, we only used Sahamyab.com/stocktwits. The users of this microblog are not all the investors of TSE. For example, the stockbrokers that have a great impact on market prices are not active on this site. So, in order to achieve more comprehensive indicators, we need to use other resources to incorporate the views of different groups of investors.

Another limitation of this work is that the amount of data used, and the number of stocks examined was limited. In order to achieve more general results about TSE, we need to consider more stocks and longer time periods. In addition, we need more data to build a larger lexicon and to develop more robust classifiers for sentiment classification.

Moreover, in this study, the calculation of the trust coefficients is static which means they are computed once and remain constant after that. Therefore, we should update them dynamically to have more informative coefficients.

In this study, we considered the relationships between the sentiment indices and the stock features as linear relations. So, we used linear regression models for prediction. However, many of these relations may be more complicated and non-linear. Therefore, in the future, we should consider non-linear models such as neural networks too.

The Persian language is a challenging language for sentiment analysis. For example, regarding the data of this study, there are lots of colloquial words in the comments and recognizing the sentence boundaries is very difficult due to the lack of punctuations. One of our most challenging tasks in the feature is to find solutions for these kinds of problems. Furthermore, one of the important tasks in sentiment analysis is considering negation, which was very difficult to address in this study due to the abovementioned characteristics of our text data. In the future, we need to come up with a solution to this issue as well.